\documentclass[letterpaper]{pac97}    
\usepackage{epsfig}    
\begin{document}

\title{ACCELERATION FOR THE $\mu^+\mu^-$ COLLIDER}
\author{D.~Summers, Dept.~of Physics, University of Mississippi--Oxford,
        University, MS 38677 USA, \\
        D.~Neuffer, Fermi National Accelerator Laboratory, Batavia, IL 60510
        USA, \\
        Q.--S.~Shu, Thomas Jefferson National Accelerator Facility,
        Newport News, VA 23606 USA, and \\
        E.~Willen, Brookhaven National Laboratory, Upton, NY 11973 USA}

\maketitle

\vspace*{-4.0cm}
\leftline{
Proceedings, \, 1997 Particle Accelerator Conference, \,
Vancouver, BC, Canada \, (12 -- 16 May 1997) 624--626}
\vspace*{2.8cm}

\begin{abstract}  
We discuss possible acceleration scenarios and methods for a
$\mu^+-\mu^-$ collider. The accelerator must take the beams from 
$\sim$100 MeV to
2 TeV within the muon lifetime 
$(2.2 \times 10^{-6} E_\mu/m_\mu \ \mu${S}),
while compressing bunches of $\sim$10$^{12}$ muons from m to 
cm bunch lengths.
Linac, recirculating linac, and very rapid-cycling synchrotron approaches are
studied. Multiple recirculating linac approaches are matched to the muon
lifetime and appear readily feasible. Rapid-cycling approaches require
innovations in magnet designs and layouts, but could be much more affordable.
\end{abstract}

\section{INTRODUCTION}

   For a $\mu^+\mu^-$ collider [1], muons must be rapidly accelerated to high
energies while minimizing the kilometers of radio frequency (RF) cavities and 
magnet bores.  Cost must be moderate. Some muons may be lost to
decay but not too many.  As the muon energy increases and the bunch length
decreases, higher frequency, higher gradient RF cavities may be used to
reduce cost.

\section{100 MeV $\rightarrow$ 2 GeV USING RF = 2 GV}

   This is the initial acceleration of cooled muons.  The bunch length
decreases from 2 m to 20 cm. A single pass 2 GV linac is used.
The RF frequency increases from 10 to 100 MHz from entrance to exit.
93\% of the muons survive. 

\section{2 GeV $\rightarrow$ 25 GeV USING RF = 2.5 GV}

This is the first recirculating ring and has 2.5 GV of 100 MHz RF [2].  A 
superconducting magnet with 10 bores, each with a different fixed field, is 
used to pass the muons through a pair of linacs 10 times.  The design is 
similar to the TJNAF in Virginia.
92\% of the muons survive.

\section{25 GeV $\rightarrow$ 250 GeV USING RF = 6 GV}

This stage uses a single ring of fast ramping $cos\,\theta$ dipoles [3].  Thin
stranded copper conductor is used at room temperature to achieve a 4 Tesla
field. The low duty cycle is exploited to keep the $I^2R$ losses reasonable.
6 GV of 350 MHz RF is distributed around the ring and accelerates the muons 
from 25 GeV to 250 GeV in 40 orbits.  
85\% of the muons survive.  

\begin{table}[!htb]
\begin{center}
\caption{Fast ramping $cos\,\theta$ dipole parameters.} 
\renewcommand{\arraystretch}{1.1}
\begin{tabular}{lc} \hline \hline
Coil inner radius  & 2 cm \\
Magnet length      & 10 m \\
Field              & 4 Tesla \\
Current            & 29.5 kA \\
Stored Energy      & 160 kJ \\
Inductance         & 370 $\mu$H \\
Coil Resistance    & 19\,000 $\mu\Omega$ \\
Ramp time, 10\% to 90\%  & 360 $\mu$S \\
Power Supply Voltage     & 31.2 kV \\
Storage Capacitance &  340 $\mu$F \\
I$^2$R magnet heat per cycle  &  9400 J \\
Magnet temperature rise per cycle & 0.13 $^0$C \\
Power into magnet @ 15 Hz   & 141 kW \\
Number of Dipoles for a ring  & 144 \\
Total power @ 15 Hz          & 20 MW \\
\hline \hline
\end{tabular}
\end{center}
\end{table}

\section{250 GeV $\rightarrow$ 2 TeV USING RF = 25 GV}

  For the final stage we consider two 2200\,m radius hybrid rings [4] of fixed
superconducting magnets alternating with iron magnets ramping at 200 Hz and
330 Hz between full negative and full positive field.  Muons are given 25 GV 
of RF energy (800 MHz) per orbit.  The RF is divided
into multiple sections as at LEP, so that magnetic fields and energies 
will match around the rings.
The first ring has 25\%
8T magnets and 75\% $\pm$2T magnets and ramps from 0.5T to 3.5T during 54 
orbits.
The second has 55\%
8T magnets and 45\% $\pm$2T magnets and ramps from 3.5T to 5.3T during 
32 orbits.  
The packing fraction is taken as 70\% in each ring.
Acceleration is from 250 GeV/c to 2400 GeV/c
and requires a total of 86 orbits in both rings; 82\% of the muons survive.

\begin{equation}
\hbox{SURVIVAL} = \prod_{N=1}^{86} \exp\left[{{-2\pi{R}\,m} \over 
{[250 + (25\,N)]\,c\tau}}\right] = 82\%
\end{equation}

  Consider the power consumption of an iron magnet which
cycles from a full -2T to a full +2T.  First calculate the energy,
$W$, stored in a 2T field in a volume 6\,m long,\, .03\,m high, and 
.08\,m wide.
$\mu_0$ is $4\pi\times 10^{-7}$.

\begin{equation}
W = {B^2\over{2{\mu_0}}}[\hbox{Volume}] = 23\,000 \ \hbox{Joules}
\end{equation}

Next given 6 turns, an LC circuit capacitor, and a 250 Hz frequency; estimate 
current, voltage, inductance, and capacitance. The height, $h$, of the 
aperture is\, .03\,m.  
The top and bottom coils may be connected as two separate
circuits to halve the switching voltage.   

\begin{equation}
B = {{\mu_0\,NI}\over{h}}  \quad\rightarrow\quad 
I = {{Bh}\over{\mu_0\,N}} = 8000 \ \hbox{Amps}
\end{equation}
\begin{equation}
W = .5\,L\,I^2  \quad\rightarrow\quad L = {2\,W\over{I^2}} = 
720\,\mu\hbox{H}
\end{equation}
\begin{equation}
f = {1\over{2\pi}}\sqrt{1\over{LC}}  \quad\rightarrow\quad
 C = {1\over{L\,(2\pi f)^2}} = 560\, \mu\hbox{F}
\end{equation}
\begin{equation}
W = .5\,C\,V^2  \quad\rightarrow\quad V = \sqrt{2W\over{C}} = 9000 \ 
\hbox{Volts}
\end{equation}

Now calculate the resistive energy loss, which over time is equal to 1/2
the loss at the maximum current of 8000 Amps.  The 1/2 comes from the 
integral of cosine squared.  
A six-turn copper conductor 3\,cm thick, 10\,cm high, 
and 7800\,cm long has an $I^2R$ power dissipation of 15 kilowatts.

\begin{equation}
R = {7800 \ (1.8\,\mu\Omega\hbox{-cm})\over{(3) \, (10)}} = 470\,\mu\Omega
\end{equation}

Now calculate the dissipation due to eddy currents in this conductor, which will
consist of transposed strands to reduce this loss [5--7].  
To get an idea, take the maximum B-field
during a cycle to be that generated by a 0.05m radius conductor carrying
24000 amps.  
This ignores fringe fields from the gap which will make the real answer higher.
The eddy current loss in a rectangular conductor made of square wires 
1/2 mm wide with a perpendicular magnetic field is as follows.  
The width of the wire is $w$.

\begin{eqnarray}
B & = & {{\mu_0\,I}\over{2\pi r}} = 0.096 \ \hbox{Tesla}  \\
P & = & \hbox{[Volume]}{{(2\pi\,f\,B\,w)^2}\over{24\rho}}  \\
  & = & [.03 \ .10 \ 78]\, {{(2\pi \ 250 \ .096 \ .0005)^2} \over 
{(24)\,1.8\times{10^{-8}}}} = 3000 \ 
\hbox {watts}
\nonumber
\end{eqnarray}

A similar calculation shows that the cooling water tube losses due to eddy
currents can be held to 1200 watts. The tubes must be made of a high 
resistivity material such as 316L stainless steel.

\begin{table}[!htb]
\begin{center}
\caption{Soft magnetic material properties [8].} 
\renewcommand{\arraystretch}{1.1}
\tabcolsep=0.6mm
\begin{tabular}{llccc} \hline \hline
                   &          &                      & B   &           \\
Material                   & Composition & $\rho$    & Max & H$_c$      \\
                   &             & $\mu\Omega$-cm & T       & Oe  \\
                                                                         \hline
Pure Iron [9]     & Fe 99.95,\, C .005      & 10              & 2.16 & .05  \\
1008 Steel         & Fe 99,\, C .08           & 12             & 2.09 &  0.8 \\
Grain--Oriented &  Si 3,\, Fe 97           & 47                & 1.95 & .1  \\
NKK Super E-Core & Si 6.5,\, Fe 93.5       & 82                & 1.8  &     \\
Supermendur  [10]  & V 2,\, Fe 49,\, Co 49   & 26              & 2.4  & .2   \\
Hiperco 27   [11]  & Co 27, Fe 71            & 19              & 2.36  & 1.7  \\
Metglas 2605SA1          & Fe 81,\, B 14,\, Si 3 &  135   &  1.6    & .03  \\
\hline \hline
\end{tabular}
\end{center}
\end{table}

   Eddy currents must be reduced in the iron not only because of the increase
in power consumption and cooling, but also because they introduce multipole
moments which destabilize beams.  If the laminations are longitudinal,
it is hard to force the magnetic field to be parallel to the laminations
near the gap.  This leads to additional eddy current gap losses [12].  
So consider a magnet with transverse laminations as sketched in Fig.~1
and calculate the eddy current losses. 
The yoke is either
0.28\,mm thick 3\% grain oriented silicon steel [13] 
or 0.025\,mm thick Metglas 2605SA1 [14, 15].
The pole tips are 0.1\,mm thick Supermendur [10] to increase 
the field in the gap [16].

\begin{eqnarray}
\lefteqn{\hbox{P(3\% Si--Fe)} 
=  \hbox{[Volume]}{{(2\pi\,f\,B\,t)^2}\over{24\rho}} = 27 \, \, \hbox{kw}}\\
& &   =  [6 \, ((.42 \ .35) - (.20 \ .23))]\, 
{{(2\pi \ 250 \ 1.6 \ .00028)^2} \over 
{(24)\,47\times{10^{-8}}}}  \nonumber 
\end{eqnarray}

Similar calculations for the eddy current losses in a Metglas yoke and in 
Supermendur pole tips yield much lower values, 75 and 210 watts, respectively.

%

\begin{figure}[htb]
\begin{center}
\mbox{\epsfig{file=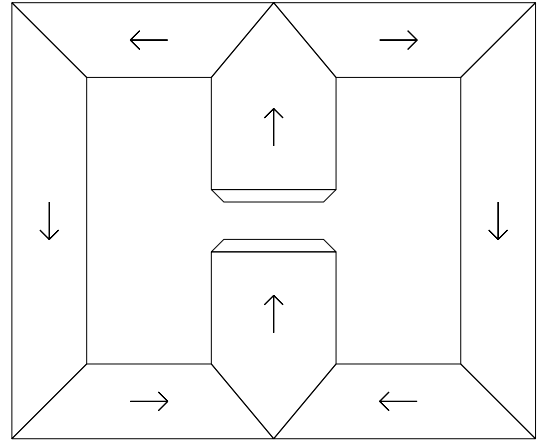,width=0.85\columnwidth}}
\caption{
H frame magnet lamination with grain 
oriented 3\%\,Si--Fe steel.  The 
arrows show both the magnetic field direction and the grain direction of 
the steel. Multiple pieces are used to exploit the high permeability and 
low hysteresis in the grain direction [17].
If Metglas 2605SA1 is used for the yoke, multiple pieces are not needed, except
for the poles.
The pole tips are an iron--cobalt 
alloy for flux concentration exceeding 2 Tesla.}
\end{center}
\end{figure}

   Eddy currents are not the only losses in the iron.  Hysteresis 
losses,
$\int{\bf{H}}{\cdot}d\,{\bf{B}}$, scale  
with the coercive force,
H$_c$, and increase linearly with frequency.
Anomalous loss [9] which is difficult to calculate
theoretically must be included.  Thus I now use functions fitted 
to experimental measurements of 0.28\,mm thick 3\% grain oriented 
silicon steel [18], 
0.025\,mm thick Metglas 2605SA1 [14],
and 0.1\,mm thick Supermendur [18].

\begin{table}[!htb]
\begin{center}
\caption{Magnet core materials.}
\renewcommand{\arraystretch}{1.1}
\tabcolsep=1.0mm
\begin{tabular}{lcccc} \hline \hline
Material            & Thickness   & Density     & Volume     & Mass \\
                    & (mm)        & (kg/m$^3$)  & (m$^3$)    & (kg) \\ \hline
3\% Si--Fe & 0.28   & 7650        & 0.6        & 4600 \\
Metglas             & 0.025       & 7320        & 0.6        & 4400 \\
Supermendur         & 0.1         & 8150        & 0.01       & 90   \\
\hline \hline
\end{tabular}
\end{center}
\end{table}

\begin{eqnarray}
\hbox{P(3\% Si--Fe)} & = & 4.38 \times 10^{-4} \, f^{1.67} \, B^{1.87}    \\   
\nonumber            & = & 4.38 \times 10^{-4} \, 250^{1.67} \, 1.6^{1.87} \\ 
\nonumber            & = & 10.7 \, \, \hbox{w/kg} = 49 \, \, \hbox{kw/magnet} 
\end{eqnarray}

\begin{eqnarray}
\hbox{P(Metglas)}    & = & 1.9 \times 10^{-4} \, f^{1.51} \, B^{1.74}    \\   
\nonumber            & = & 1.9 \times 10^{-4} \, 250^{1.51} \, 1.6^{1.74} \\ 
\nonumber            & = & 1.8 \, \, \hbox{w/kg} = 7.9 \, \, \hbox{kw/magnet}
\end{eqnarray}

\begin{eqnarray}
\hbox{P(Supermendur)} & = & 5.64 \times 10^{-3} \, f^{1.27} \, B^{1.36}    \\   
\nonumber            & = & 5.64 \times 10^{-3} \, 250^{1.27} \, 2.2^{1.36} \\ 
\nonumber            & = & 18 \, \, \hbox{w/kg} = 1.6 \,\, \hbox{kw/magnet}
\end{eqnarray}

\begin{table}[!hbt]
\begin{center}
\caption{Power consumption for a 250 Hz dipole magnet.}
\renewcommand{\arraystretch}{1.1}
\begin{tabular}{lcc} \hline \hline
Material                   &    3\% Si--Fe     &       Metglas    \\       
Coil Resistive Loss    & 15\,000 watts &      15\,000 watts \\
Coil Eddy Current Loss &  4200 watts      &      4200 watts  \\
Total Core Loss             &  50\,600 watts        &      9500 watts \\ \hline
Total Loss                  &  69\,800 watts &  28\,700 watts  \\ \hline \hline
\end{tabular}
\end{center}
\end{table}

   In summary, a 250 Hz dipole magnet close to 2 Tesla looks possible as long
as the field volume is limited and one is
willing to deal with stranded copper and thin, low hysteresis laminations.
Total losses can be held to twice the I$^2$R loss in the copper alone, 
using Metglas.

  The 1925 ramping dipoles which are required consume 56 megawatts when running.
Given a 15 Hz refresh rate for
the final muon storage ring [1], 
the average duty cycle for the 250 $\rightarrow$
2400 GeV/c acceleration rings is 6\%. So the power falls to 4 megawatts,
which is small.

\begin{table}[!htb]
\begin{center}
\caption{4 acceleration stages; 60\% $\mu^{\pm}$ survival overall.} 
\renewcommand{\arraystretch}{1.1}
\tabcolsep=0.7mm
\begin{tabular}{lcccc} \hline \hline
      & Linac  & 10 Bore    & Cu $cos\,\theta$  &  Iron \& SC    \\
      &        & SC Magnet  & Magnet      &  Magnets       \\ \hline
E(GeV) & 0.1$\rightarrow$2 & 2$\rightarrow$25 & 25$\rightarrow$250 & 
                                                250$\rightarrow$2000 \\
RF(MHz) & 10$\rightarrow$100 & 100 & 350 & 800 \\
N(turns) & 1  & 10  & 40  & 86 \\
RF(GV)   & 2  & 2.5 & 6   & 25 \\
Length(km) & 0.4  & 0.36  & 2.3 & 14 \\
$\tau$(ms) & 0.0013 & 0.012 & 0.307 & 4.0 \\
$\mu$ Bunch(cm) & 200$\rightarrow$20 & 6 & 2 & 1 \\
$\mu$ Survival  & 93\% & 92\% & 85\% & 82\% \\
\hline \hline
\end{tabular}
\end{center}
\end{table}

\end{document}